\documentclass[letterpaper]{JHEP3}
\usepackage{epsfig}

\usepackage{amssymb}

\usepackage{amsfonts}
\usepackage{epsfig}
\usepackage{latexsym}
\usepackage{graphicx}
\def\bequ{\begin{equation}}
\def\eequ{\end{equation}}
\def\barr{\begin{array}}
\def\earr{\end{array}}
\def\half{{1\over 2}}
\def\ben{\begin{equation}}
\def\een{\end{equation}}
\def\bena{\begin{eqnarray}}
\def\eena{\end{eqnarray}}

\def\bea{\begin{eqnarray}}
\def\eea{\end{eqnarray}}
\def\be{\begin{equation}}
\def\ee{\end{equation}}
\def\half{{1\over 2}}

\def\del{\partial}
\def\Tr {\mbox{Tr}}

\preprint{UCLA/04/TEP/06\\
UOSTP-04101\phantom{abcd}\\
OU-HET 470\phantom{abcde}\\
{\tt hep-th/0403249}\phantom{ab}}

\title{Dilatonic Repulsons  and Confinement
via the AdS/CFT
Correspondence
}

\author{Dongsu Bak$^1$, Michael  Gutperle$^2$,
Shinji Hirano$^3$ and Nobuyoshi Ohta$^4$ \\
\vskip 8mm
\centerline{
$^1$Physics Department, University of Seoul, Seoul 130-743, Korea}
\vskip 3mm
\centerline{$^2$
Department of Physics and Astronomy, UCLA, Los Angeles,
CA 90095, USA}
\vskip 3mm
\centerline{$^3$ Department of Physics, Technion, Israel Institute of
  Technology, Haifa 32000, Israel}
\vskip 3mm
\centerline{$^4$ Department of Physics, Osaka University, Toyonaka,
  Osaka 560-0043, Japan}\\
\vskip 8mm
\email{dsbak@mach.uos.ac.kr}\ ,\hspace{0.1cm}
\email{gutperle@physics.ucla.edu}\ , \\
\email{hirano@physics.technion.ac.il}\ ,\hspace{0.1cm}
\email{ohta@phys.sci.osaka-u.ac.jp}}

\abstract{We study a class of dilatonic deformations of asymptotically
AdS$_5\times S^5$ geometry analytically and numerically.
The spacetime is non-supersymmetric and suffers from
a naked singularity. We propose that the causality bound may
serve as a criterion for such a geometry with a naked
singularity to still make sense in the AdS/CFT correspondence.
We show that the static string,
the one corresponding to a large Wilson loop in the dual gauge theory,
reveals confinement in a certain range of parameters of our solutions,
where the singularity exhibits the repulsion that can well cloak the
singularity from the static string probe.
In particular, we find the exact expression for the tension of
the QCD strings.
We also discuss a possible interpretation of our solution in terms of
unstable branes and their tachyon matter.}

\begin{document}
\baselineskip16pt
\parskip=4pt

\tableofcontents
\section{Introduction}

The AdS/CFT correspondence~\cite{Maldacena:1997re,Gubser:1998bc,Witten:1998qj}
firmly established a duality between gravity and gauge theory, providing
a remarkable realization of holography~\cite{'tHooft:gx,Susskind:1994vu}.
There has been a plethora of applications and generalizations thus
far, yet the non-supersymmetric cases have been less studied.
A simple way to obtain the non-supersymmetric generalization is to
add deformations that break supersymmetry.
One of the simplest such deformations is to turn on only the
dilaton. Such a dilatonic deformation on the one hand typically breaks
all the supersymmetries and often suffers from a naked singularity. On the
other hand it is simple and therefore quite tractable.

A non-singular dilatonic deformation, the ``Janus'' solution, was
found in \cite{Bak:2003jk}, which turns out to be dual to a
non-supersymmetric dilation invariant deformation of
${\cal N}=4$ Supersymmetric Yang-Mills theory (SYM)
by a certain exactly marginal operator.
Even further the non-perturbative stability of the ``Janus'' solution
was established against a broad class of fluctuation modes by a
compelling argument developed in \cite{Freedman:2003ax}.
Thus the ``Janus'' solution was proven to be on a firm footing in the
AdS/CFT correspondence, despite the lack of supersymmetry.

With this success, though moderate, it may be worthwhile to further
explore possible dilatonic deformations of different characteristics.
In fact there exists a wide class of simple dilatonic deformations
(see, for example,~\cite{Gubser:1999pk,Kehagias:1999tr,Nojiri:1999sb}).
It seems, however, rather rare that the dilatonic deformations lead to
non-singular geometries. In general relativity a geometry with a naked
singularity should be dismissed because of cosmic censorship. In
string theory however a naked singularity can make sense if
a brane interpretation is given, or a stringy resolution exists.
In terms of the AdS/CFT correspondence, the problem of the singular
geometry would be manifested by a causal violation of holography
\cite{Kleban:2001nh}, -- for the bulk/boundary correspondence to make
sense, it is prohibited that any signal, sent from a point P on the
boundary, reaches a distant point Q on the boundary, faster by
traveling through the bulk than by traveling along the boundary.
For instance, the negative mass AdS-Schwarzschild black hole is a
typical example of geometry with a naked singularity.
Indeed the causality bound of \cite{Kleban:2001nh} excludes it from
being a sensible geometry in the AdS/CFT correspondence.

In this paper, we take the causality bound as a postulate to diagnose
which of singular geometries with asymptotically AdS space may make
sense in string theory.
The dilatonic deformations we consider in this paper all suffer from
a (timelike) naked singularity.
In fact adopting the causality bound rules out a certain range of
parameter space of our solutions, but leaves a wide range to be
sensible as gravity dual of the boundary gauge field theory.
Physically the causality condition imposes the lower (non-negative)
bound on the mass of our singular geometry.
In other words, sufficiently energetic dilatonic deformations, even
with the naked singularity, can clear the causality bound.

In particular three types of deformations are studied; the
global patch, Poincar\'e patch, and hyperbolic slicing.
In all cases, we have three independent parameters; $\phi_\infty$, the
constant part of the dilaton, $k$, the strength of the nontrivial
profile of the dilaton, and $A$ (or $\mu$), the energy of the geometry.
The causality bound sets the lower (non-negative) bound on $A$ (or
$\mu$) for any given $k$.
In the case of the global patch, we carry out a numerical analysis to
solve the condition for the causality bound and find the positive
lower bound on $A$ for any given $k$. In the case of the Poincar\'e
patch, the causality bound is somewhat trivial and analytically
solved. It simply imposes $\mu$ to be non-negative.
Our claim is that, within these parameter ranges, the AdS/CFT
correspondence makes sense, and our geometry is dual to some
(non-supersymmetric) states, labeled by $k$ and $A$ (or $\mu$), in
${\cal N}=4$ SYM, where $k$ and $A$ (or $\mu$) are the expectation
values of (supersymmetric completion of) $\Tr\, F^2$ and the energy
$T_{00}$ respectively, following~\cite{Balasubramanian:1998sn,Banks:1998dd}.

There is yet another distinct range of parameters for our
solutions. In this range, the naked singularity exhibits a strong
repulsion overwhelming the attraction due to the negative cosmological
constant. This can be regarded as an example of the {\it repulson}
geometry~\cite{Behrndt:1995tr,Kallosh:1995yz,Cvetic:1995mx}.
A typical case of the repulson is again the negative mass black hole.
We, however, find the overlap of the parameter ranges for
the repulson and the causality. Thus the repulsion occurs
even when the mass is positive in our case.

The repulson is, as we argue, a potential indication of confinement in
the dual gauge theory. We demonstrate our claim in the case of the
Poincar\'e patch, by calculating the static string, corresponding to
the Wilson loop in the dual gauge theory~\cite{Rey:1998ik,Maldacena:1998im}.
For a small loop, the static string is not pulled down deep enough to
feel the deformation of geometry. Thus the quark-antiquark potential
is shown to be Coulombic as in the AdS case. However, as the loop gets
larger and larger, we find a (confining) scale far beyond which the
quark-antiquark potential becomes linear.
Unlike the AdS case, the long string would not be dragged far deep
according to the simple IR/UV relation \cite{Peet:1998wn}, but stops
penetrating beyond the confining scale.
In other words, the repulsion is strong enough to
cloak the naked singularity from the static string. The parameters
can be easily adjusted to make the confining scale be far from the
singularity, thus rendering our analysis reliable.
However, we note a caveat that there is another branch for the
solution of the static string dynamics, wherein the Wilson loop does
not show the confining behavior and the long string is drooped
all the way down to the singularity. It is not clear how to interpret
this branch or ``phase'' in terms of the renormalization group
flow of the dual gauge theory, since the regular branch alone covers
all the length scale of the dual gauge theory.

Finally it is natural to ask if there is any brane interpretation for
our geometry. On this score, it may be tempting to identify the
singularity of the global deformation as a collection of
unstable D0 branes smeared over $S^5$. The reasoning follows from
the fact that the singularity is pointlike and does not carry any
Neveu-Schwarz (NS) or Ramond-Ramond (RR)  charge, but sources the
dilaton and gravity. A speculation on this point will be elaborated
and expanded in the discussion.

The organization of our paper is as follows. In section 2, we set up
the ansatz for our dilatonic deformations,
and propose its interpretation in terms of the AdS/CFT
correspondence. In section 3, we discuss the case of deformation
in the global coordinates. Particular emphasis is put on the
causality bound and the condition for the repulsion.
For the most part, we solve the problem numerically and find
the causality (lower) bound on the mass for any given $k$, whereas
the condition for the repulsion is found exactly.
Sections 4, 5 and 6 deal with the case of deformation in the Poincar\'e
patch. We find the exact analytic solution and the simple causality bound.
The condition for the repulsion turns out to be the same as that in
the global coordinates. We show evidence for the confinement by
calculating the Wilson loop (static string), and clarify its relation
to the repulsion. We also compute the tension of the QCD string exactly.
Final section is devoted to discussions, and an appendix is provided
for the case of deformation in the hyperbolic slicing.

\section{Preliminary}\label{sec2}

We consider the following simple non-supersymmetric dilatonic
deformations of AdS$_5\times S^5$:
\bea
ds^2&=&l^2\left[
-h(r)g(r)dt^2+{h(r)\over g(r)}dr^2+r^2ds_{3,K}^2+d\Omega_5^2
\right]\,\label{metric}\\
\phi&=&\phi(r)\ ,\label{dilaton}\\
F_5&=&Q l^5\left(d\Omega_5+\ast d\Omega_5\right)\ ,
\label{5form}
\eea
where
$d\Omega_5$ denotes the volume form of the
five sphere and
$l$ is the radius of undeformed AdS$_5$ and $S^5$.
The subscript $K$ in $ds_{3,K}^2$ labels the curvature of the maximally
symmetric $3$-dimensional space, and $Q$ is a constant.
In our convention, the curvature $K$ is equal to $+6$ for sphere and
$-6$ for hyperboloid.
The isometry of this geometry in general case is an
$SO(2)\times G_3\times SO(6)$ subgroup of $SO(2,4)\times SO(6)$,
where $G_3$ is the isometry group of the maximally symmetric
$3$-dimensional space with curvature $K$, i.e.
$SO(4)$ for $K>0$, $ISO(3)$ for $K=0$, and $SO(1,3)$ for $K<0$.
The $SO(2)$ is the time translation symmetry whereas the $SO(6)$
is the isometry of $S^5$.

The equations of motion for the type IIB string are given by
\bea
\nabla^2\phi&=&0\ ,\nonumber\\
d\ast\! F_5&=&0\ ,\nonumber\\
R_{MN}&=&\half\del_M\phi\del_N\phi+{g_s^2\over 96}
F_{MPQRS}F_N^{\hspace{0.2cm}PQRS}\ ,
\eea
together with the Bianchi identity $dF_5=0$.
The above ansatz leads to
\bea
\phi'&=& {k/l^3\over  r \psi}\ , \label{eqone}\\
\left(\ln h\right)'&=&{(k/l^3)^2\over 6r\psi^2}\ ,
\label{hhh} \\
\psi'&=&r\left((g_sQ l/2)^2r^2+K/3\right)h\ ,
\label{iii}
\eea
where we have introduced $\psi=r^2g$ and $k$ is a constant.
Thus the Einstein equations boil down to\footnote{The $K=+6$ case was
previously discussed in \cite{Das:2001rk}.}
\be
r\left(\ln\psi'\right)'=
\frac{6r^2+{K\over 6}}{2r^2+{K\over 6}}
+{(k/l^3)^2\over 6\psi^2}\ .
\label{key}
\ee
Here the constant $Q$ for the $5$-form flux has been chosen to be
$4/(g_s l)$ so that the value of the flux, in our normalization, is
$(2\pi\sqrt{\alpha'})^4N$ with $l^4=4\pi g_s N\alpha'^2$.

In general the deformation $\delta\varphi$ of supergravity fields
behaves asymptotically as
\begin{equation}
\delta\varphi= a_\Delta r^{\Delta-4}+ b_\Delta
r^{-\Delta}+ O(r^{-\Delta-1}) \ ,
\end{equation}
where the non-normalizable mode $a_\Delta$ corresponds, in the dual
CFT, to the source \cite{Gubser:1998bc,Witten:1998qj}, whereas
the normalizable mode $b_\Delta$ to the expectation value of the
operator ${\cal O}_{\Delta}$
of dimension $\Delta$~\cite{Balasubramanian:1998sn,Banks:1998dd}.
Here we are only interested in dimension four operators,
since we are turning on only the dilaton which back reacts to the
metric and thus we are deforming only massless modes.
The dilaton corresponds to the supersymmetric completion of $\Tr\, F^2$,
i.e. the ${\cal N}=4$ SYM Lagrangian density
${\cal L}_{CFT}$ itself.
In our case, the dilaton behaves asymptotically as
\begin{equation}
\phi= \phi_\infty- {k/l^3\over 4} r^{-4}+ O(r^{-5})\ ,
\end{equation}
where $\phi_\infty$ is not related to $k$ at all, as might be expected for
the singular geometry.
This translates, in the dual gauge theory, to the shift of the
coupling constant
\begin{equation}
{\cal L}_{deformed}= (1-\phi_\infty){\cal L}_{CFT}\ ,
\label{Ldef}
\end{equation}
and to the selection of a particular state with
respect to which the operator ${\cal L}_{CFT}$ acquires the
expectation value
\begin{equation}
\langle {\cal L}_{CFT} \rangle ={k/l^3\over 4}\ .
\label{expec}
\end{equation}
The non-vanishing $k$ in general breaks all the supersymmetries.
Furthermore, as we will see later, our deformation of metric in
general leads to the non-vanishing expectation value for the energy
\be
\langle T_{00} \rangle \ne 0\ ,
\ee
where its value is given by yet another parameter $A$ (or $\mu$) in
our solutions.

Since the non-normalizable mode $\phi_\infty$ of the dilaton
corresponds to a trivial shift of the coupling constant,
the Lagrangian is essentially unaffected by the deformation. Hence
$\phi_\infty$ does not play any relevant role.\footnote{Thus we may
choose a convenient value of $\phi_\infty$. In addition we shall
set $l=1$ for the notational simplicity.} We only turn on the expectation
value for the corresponding operators, which may be interpreted as
a choice of the background, or the state, of the field theory.
The (non-supersymmetric) state thus chosen is basically characterized by two
parameters $k$ and $A$ (or $\mu$).

Note also that the bosonic part of $\langle{\cal L}_{CFT} \rangle$ is
given by
\begin{equation}
\langle {\cal L}_{CFT}\rangle={1\over 2 g^2_{YM} }\langle \Tr(E^2-B^2)
\rangle={k/l^3\over 4}\ .
\end{equation}
Thus one may call the $k > 0$ ($k < 0$) case
electric (magnetic) where the electric (magnetic) contribution wins
over the magnetic (electric) one.

In the following, we shall discuss the case of $K=+6$ (global coordinates)
and $K=0$ (Poincar\'e patch) in detail.
We relegate the $K=-6$ case (hyperbolic slicing) to the Appendix.

\section{Global Deformation}
\label{sec3}

In this section we analyze the dilatonic deformation in the
global coordinates of AdS. This is the case of $K=+6$ in
Section~\ref{sec2}. The dual boundary field theory lives on $R\times S^3$.
The dilatonic deformation (\ref{metric})-(\ref{5form})
corresponds to turning on constant expectation values of the Lagrangian
density ${\cal L}_{CFT}$ as well as, as we will see below, of the energy
$T_{00}$. We would like to emphasize that the state with such
expectation values surely exists within ${\cal N}=4$ SYM. Our ansatz
(\ref{metric})-(\ref{5form}) is the most general one that respects
the properties of such state in the field theory.
However, as we will show below, our proposed
dual gravity solution inevitably contains a timelike naked singularity.
Instead of giving up the singular solution,
we will utilize the AdS/CFT correspondence as a diagnostic tool to
determine which of singular solutions may be acceptable.

The equations of motion (\ref{eqone})-(\ref{iii}) cannot
be solved analytically in this case. However,
the asymptotic expansion of the solution around $r=\infty$ (boundary)
and $r=0$ (singularity) can be found.
The expansion around $r=\infty$ reads
\bea
\psi&=& h_\infty\left( r^4 +r^2  - A\right)  +
{k^2\over h_\infty}\left( {1\over 48\, r^4} - {11\over 720\, r^6}\right) +
O\left({1\over r^8}\right)\ , \label{expone}\\
\phi&=&\phi_\infty - {k\over h_\infty}
\left( {1\over 4\, r^4} - {1\over 6\, r^6} +
{A+1\over 8\, r^8}\right)+
O\left({1\over r^{10}}
\right)\ ,\\
h&=&h_\infty - {k^2\over h_\infty} \left(
{1\over 48\, r^8} - {1\over 30\, r^{10}} \right)
+ O\left({1\over r^{12}}\right)\ .
\eea
It is easy to see that the overall factor
$h_\infty$ in (\ref{expone}) can
be absorbed by the rescaling of time in the metric
(\ref{metric}).  Hence one can set, without loss of generality,
$h_\infty=1$, and the solution depends on the three integration
constants $k, \phi_\infty$ and $A$.
The asymptotic expansion provides
the initial conditions for the numerical study
of the equations of motion.

The behavior of the solution near $r=0$ is given by the following expansion:
\bea
\psi&=& \psi_0\left(1  + {2 h_0
\over 2\psi_0
+ {k^2\over 6 \psi_0} }r^{2+{k^2\over 6 \psi_0^2}}\right) +
O\left( r^{4+{k^2\over 6 \psi_0^2}},\,\,
r^{2+{k^2\over 3\psi_0^2}}\right)\ ,\label{expthree}\\
h&=& h_0\left( r^{k^2\over 6 \psi_0^2}- {24  h_0 k^2
\over \psi_0^3 \left(12+ {k^2\over \psi_0^2}\right)^2}\, r^{2+{k^2\over 3
\psi_0^2}}\right) + O\left( r^{4+{k^2\over 3\psi_0^2}}\right)\ ,
\label{expfour}\\
\phi&=&\phi_0 + {k\over  \psi_0} \ln r +O\left( r^{2+{k^2\over 6
 \psi_0^2}}\right)\ .
\eea
For the solution with $k\neq 0$, one finds that there is always a naked
(curvature) singularity at $r=0$. In addition the dilaton diverges; for
$k>0$ ($k<0$), the string coupling goes to zero (infinity) at $r=0$.
Note that one can set $h_\infty=1$ by the rescaling of time, but
it is in general not possible to set both of $h_\infty$ and $h_0$ to
one. One can prove, given $h_\infty >0$,
the function $\psi(r)$ increases monotonically
with $\psi_0 > 0$ for any nonvanishing $k$.

We will solve the equations of motion
numerically with the initial conditions, given by $\phi_\infty, k$
and $A$, at some large cutoff $r_f$.

\subsection{The causality bound}
\label{sec3.1}

We have seen that there is a three parameter family of solutions.
They all have the timelike naked singularity, but are the only
possible candidates of the dual of the above-mentioned state in
${\cal N}=4$ SYM.
However, in order for them to be sensible geometries in the AdS/CFT
correspondence, they have to satisfy the following causality condition:
No information, sent from a point P to Q on the
boundary, can propagate faster through the bulk than along the
boundary. This is a causal consistency of holography in the bulk/boundary
correspondence.
In our particular case, the causality bound is found to be
\cite{Kleban:2001nh}
\begin{equation}
\pi \
\le \  2\int_0^\infty dr{ r^2 \over \psi(r)} \,.
\label{cau}
\end{equation}
A typical singular spacetime that violates the causality bound
(\ref{cau}) is the negative mass AdS-Schwarzschild black hole, as is
easily seen from,
\begin{equation}
\pi= 2 \int_0^\infty  {dr\over r^2+1}\
> \ 2 \int_0^\infty {dr\over r^2+1 -A/r^2} \ ,
\end{equation}
when $A$ is negative.
Thus it is excluded from the AdS/CFT correspondence, as it should be.

In our application too, the negative $A$ is excluded. We can first
show that  $\psi(r;A) > \bar\psi(r;A)\equiv r^4+r^2-A$ as follows:
Let $\psi'=\bar\psi' \chi$. Then the equation (\ref{key}) becomes
$$
r (\ln\chi)'= {k^2\over 6\psi^2}\ ,
$$
which is solved as
\begin{equation}
\chi= e^{-\int^\infty_r dr {k^2\over 6 r\psi^2}}\ .
\label{int_a}
\end{equation}
Thus we obtain
\begin{equation}
\psi=\bar\psi+ \int_r^\infty dr\, \bar\psi'\,
(1-e^{-\int^\infty_r dr'{k^2\over 6 r'\psi^2}})\ .
\label{int_b}
\end{equation}
This proves that $\psi(r;A) > \bar\psi(r;A)$.
For $A < 0$, this reads
\begin{equation}
\pi\
> \ 2 \int_0^\infty {dr\over r^2+1 + |A|/r^2}
\
> \  2 \int_0^\infty dr{r^2\over \psi(r;A)}
\,.
\end{equation}
Hence the causality condition is violated for $A < 0$.

\begin{figure}[htb]
\vskip .5cm
\epsfxsize=2.8in
\centerline{
\epsffile{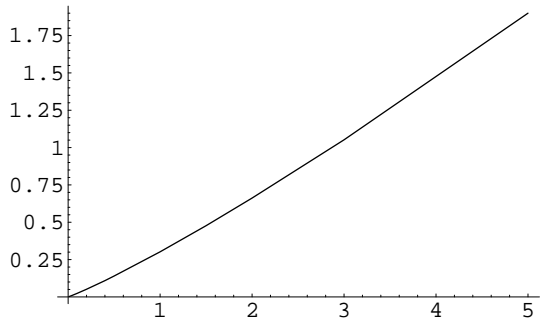}
}
\vspace{.1in}
\caption{\small Critical ADM mass parameter $A(|k|)$ as a function of $|k|$.}
\label{crit}
\end{figure}

However, we find by a numerical study that the causality bound can
actually be obeyed for a wide range of parameters $(A,k)$ with
$A\ge 0$.
The saturation of the bound provides a boundary curve $A(k)$
 in the
$(A,k)$ space, as shown in Figure~\ref{crit}. The causality condition
is satisfied when $A\ge A(k)\ge 0$ with $A(k)$ being even in $k$.

The causality bound is, strictly speaking, a necessary condition for
our spacetime to make sense in the AdS/CFT correspondence.
But once the causality bound is cleared,
we do not see any apparent obstruction for identifying our
singular spacetime with the gravity dual of the above-mentioned state in
the field theory, and we claim that our singular spacetime
with $A\ge A(k)$ is indeed acceptable in the AdS/CFT correspondence.

\subsection{The repulson}
\label{sec3.2}

We have argued that our solution, albeit doomed to have the naked
singularity, is still a viable gravity dual of a well-defined state in
the field theory, as long as the causality condition $A\ge A(k)$ is
met.
Given this status, we now proceed to discuss physical properties of
our singular geometry. The massive particle is a standard probe to study
the characteristics of the geometry. In particular we consider the
spherically symmetric geodesics of the massive particle, and will work
in the string metric. The geodesic equations,
\begin{equation}
{d^2 x^\alpha\over d\tau^2}+ \Gamma^\alpha_{\beta\gamma}{dx^\beta\over d\tau}
{dx^\gamma\over d\tau} =0\ ,
\end{equation}
reduce to the following first order equations,
\begin{equation}
{dt\over d\tau} = {C r^2 \over h \psi  e^{\phi\over 2}}\ ,
\end{equation}
with $C$ being a positive integration constant, and
\begin{equation}
\left({dr\over d\tau}\right)^2+ V(r)=0\ ,
\end{equation}
where
\begin{equation}
 V(r)={\psi \over  r^2h e^{\phi\over 2}}
\left(1- {C^2 r^2 \over \psi  he^{\phi\over 2}
}\right)\ .
\label{pot}
\end{equation}
This is a system of a zero-energy particle moving in one dimension
under the potential~(\ref{pot}).
At small $r$, the potential behaves like
\begin{equation}
V(r)={ \psi_0\over
 h_0 r^{{1\over 6}\left({k\over \psi_0}+{3\over 2}
\right)^2+{13\over 8} } } \left(1- {C^2\, r^{{19\over 8}-{
1\over 6}\left({k\over  \psi_0}+{3\over 2}
\right)^2}
\over h_0 \psi_0}\right)\,,
\end{equation}
where $ \psi_0$ and $h_0$ were defined in (\ref{expthree}) and (\ref{expfour}).

The condition for the hard-core repulsion near $r=0$ is then given by
\begin{equation}
-{\sqrt{57}+3\over 2}(\simeq -5.27) \ \le\  {k\over  \psi_0}
\ \le\  {\sqrt{57}-3\over 2}(\simeq 2.27)\ .
\label{repcon}
\ee
Outside this range the naked singularity is attractive. Hence the
quantity ${k/ \psi_0}$
is a measure of whether the singularity is attractive or repulsive.
In Figure~\ref{kpsi}, we plot
${k/ \psi_0}$ for the critical values of $A(k)$ and $k$.
\begin{figure}[htb]
\vskip .5cm
\epsfxsize=2.8in
\centerline{
\epsffile{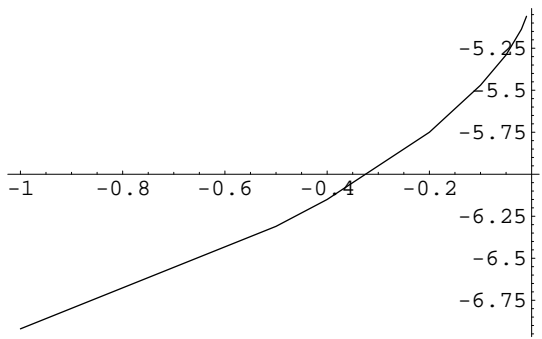}
}
\vspace{.1in}
\caption{\small $k/\psi_0$ as a function of $k$ for the critical $A(k)$.}
\label{kpsi}
\end{figure}
It is easy to see that
$k/\psi_0$ monotonically decreases, as $k$ decreases, for $k<0$,
whereas there is no value of $(A(k), k)$
which satisfies the bound (\ref{repcon}), for $k>0$.
Hence, only for $0\,\,>\,\,k\,\,>-0.02$, the singularity can be
repulsive, while saturating the causality bound.

\subsection{The ADM mass}
\label{sec3.3}

The parameter $A$ in our solution can be interpreted as the mass of
the geometry. Indeed,
using the counter term methods, one can compute the
energy momentum tensor defined by~\cite{Bala,Empa}
\begin{equation}
T_{ab}\equiv K_{ab}-
\gamma_{ab} K - 3 \gamma_{ab}
-{1\over 2}\left(R_{ab}(\gamma) -{1\over 2}\gamma_{ab}
R(\gamma)\right)
\,,
\label{masses}
\end{equation}
where $\gamma_{ab}$ is the induced metric on the boundary
and $K_{ab}$ is the extrinsic curvature of the boundary.
In our case, the mass is evaluated as
\begin{equation}
M={3\pi A\over 8 G_5}+ {3\pi\over 32 G_5}
\,,
\end{equation}
where the second term represents the contribution of the Casimir
energy on the sphere. Thus the mass for the excitation of our interest
is given by
\begin{equation}
M_{ex}={3\pi A\over 8 G_5}
\,.
\end{equation}
Restoring $l$ dependence, one has
\begin{equation}
{l^3\over G_5}= {2 N^2\over \pi}
\,,
\end{equation}
where $N$ denotes the number of D3 branes. For instance,
the Casimir energy~\cite{Bala}
is given by
\begin{equation}
E_c= {3\pi l^2\over 32 G_5}= {3 N^2 \over 16 l}\,.
\end{equation}

\subsection{Repulson vs confinement}
\label{sec3.4}

We would now like to argue that the infinitely strong repulsion
discussed above might signal the confinement in the dual field theory.
If one uses the static string probe corresponding to the Wilson loop,
the hard-core repulsion near the singularity will keep the string away
from the singularity for any given boundary separation of the
quark-antiquark pair (see Figure \ref{fig4}).
This may be expected on the basis of energetics, -- this configuration
minimizes the energy of the string.
\begin{figure}[htb]
\epsfxsize=3.8in
\centerline{
\epsffile{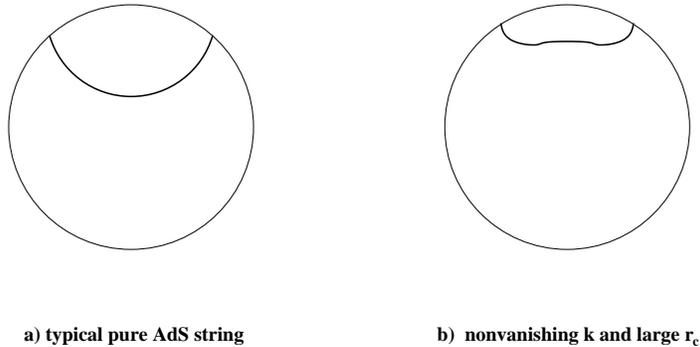}
}
\vspace{.1in}
\caption{\small String configurations in the global deformation.}
\label{fig4}
\end{figure}

Let us introduce the scale $r_c$ around which one sees an appreciable
deviation of the geometry from the pure AdS case.
We find it convenient to define the scale $r_c(<\infty)$ by
$\psi(r_c)/\psi_{AdS}(r_c)=1$
with $\psi_{AdS}(r)=r^4+r^2$.\footnote{The function $\psi(r)$
approaches asymptotically to $r^4+r^2-A(|k|)$. Thus, for large but
finite $r$, $\psi(r)/\psi_{AdS}(r)$ is less than one. Both $\psi(r)$
and $\psi_{AdS}(r)$ decrease as $r$ decreases, but $\psi(r)$ decreases
slower than $\psi_{AdS}(r)$. Thus the point where the ratio becomes
one is characteristic of the deformation.}
This defines the scale at which the deviation is of order one.
The scale $r_c$ is plotted in
Figure \ref{fig3}
as a function of $|k|$ for the critical mass parameter $A(|k|)$.
\begin{figure}[htb]
\epsfxsize=2.8in
\centerline{
\epsffile{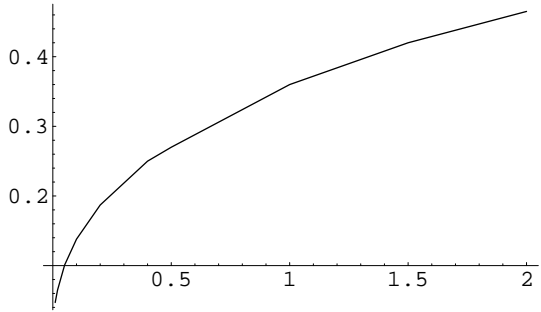}
}
\vspace{.1in}
\caption{\small $r_c$ as a function of $|k|$ for $A(|k|)$.}
\label{fig3}
\end{figure}
When the potential is repulsive, $r_c$ would be roughly the scale at
which the string turns around. As will be demonstrated below
explicitly in the case of the Poincar\'e patch,
when the separation of the quark-antiquark pair becomes larger and
larger, more and more portion of the string stays around this scale.
Hence, when the confinement occurs,
it would happen below the energy scale,
\be
E_{conf} \sim r_c\,,
\ee
via the UV/IR relation of the AdS/CFT
correspondence.\footnote{The UV/IR relation does also deviate from
the pure AdS case. Thus the estimate of the confining scale is
rough but within the accuracy of order one. As long as $r_c$ is much
larger than the AdS radius, the estimate is reasonable.}
In other words,
when the quark-antiquark separation becomes larger than $1/r_c$, the
Wilson loop would show the area law. The Faraday flux
between quark and antiquark at the boundary looks string-like with
the size of its cross section being $1/r_c$. All of these
are a consequence of the repulsive force of our geometry.
We illustrate the shape of string in Figure \ref{fig4} as a
cartoon, in the case when the potential is repulsive and
$r_c$ is much larger than the AdS radius $l$.

However, our previous numerical result shows that
for the repulsion and causality we have a rather narrow range,
$0\,>\,k \,>\,-0.02$, which reads $r_c\,\, <\,\, 0.065$.
It then implies that, to see the confining behavior,
the boundary separation must be much larger than $(1/0.065)\,l$,
that is larger than the size of the AdS space and thus is not possible.
Hence, in the global deformation, the confinement
may not be seen clearly.

The study of the static string in Section \ref{sec6} shows that the
repulsion condition is actually relaxed to $k/\psi_0 < -6$,
as compared to the massive particle case.
Since the condition is by nature local in the bulk geometry,
one might expect that the same may be true for the global deformation.
However, even with this relaxed condition, $r_c$ is still much smaller
than one. Thus the conclusion concerning the confinement
seems still negative. The conclusive answer would require a detailed
numerical study of the behavior of the string connecting quark and
antiquark in the global coordinates.

\section{Deformation in the Poincar\'e Patch}
\label{sec4}

In this section we consider a
deformation of AdS$_5\times S^5$ in the Poincar\'e patch,
thus keeping the symmetry of
$SO(2)\times ISO(3)\times SO(6) \subset SO(2,4)\times SO(6)$.
The $SO(2)$ is again the time translation symmetry, and the $ISO(3)$
is the Poincar\'e symmetry of the Euclidean $3$-dimensional plane.
This case corresponds to $K=0$ in our ansatz (\ref{metric})-(\ref{5form}).
It turns out that we can analytically obtain the exact solution
in this case.
In particular the exact analytic soluion allows us to show
evidence for the confinement in the dual gauge theory, as we will see
in Section \ref{sec6}. The dual ${\cal N}=4$ SYM theory
lives on $3+1$ dimensional flat space in this case.
As discussed in Section \ref{sec2}, the field theory background
is nontrivial, on which the Lagrangian density ${\cal L}_{CFT}$ and
the energy $T_{00}$ acquire constant expectation values.
The backgrounds parameterized by different values of $k$ and $\mu$ are
separated by the super-selection, and
it costs an infinite amount of energy to interpolate them.

Our main equation (\ref{key}) takes a particularly simple form,
\be
r(\ln\psi')'= 3 + {k^2\over 6\psi^2}\ ,
\ee
where we set the radius $l$ to unity.
The AdS boundary is located at $r=\infty$ and we require
the asymptotically AdS condition as
\begin{equation}
r^{-4} \psi \ \ \rightarrow\ \  1,
\end{equation}
as $r  \  \rightarrow\   \infty$.
The function $\psi'$ is positive definite and its magnitude decreases
monotonically as $r$ decreases.
Moreover one can show that $\psi$ cannot become zero anywhere.
Thus $\psi$ is convex, monotonically decreasing and remains positive
definite, as $r$ decreases.

Indeed one can find the exact solution, which is given implicitly by
\begin{equation}
r= (y-b)^{1-a\over 8}  (y+b)^{1+a\over 8},
\end{equation}
where
\begin{equation}
a\equiv \left(1+{k^2\over 6\mu^2}\right)^{-{1\over 2}}\,,\ \ \
b\equiv {\mu\over 2}\left(1+{k^2\over 6\mu^2}\right)^{1\over
  2}\,. \label{abrelation}
\end{equation}
with $y=\psi+ab$. Using (\ref{eqone})-(\ref{iii}), we get
\begin{equation}
h={1\over y-ab} (y-b)^{1+a\over 2}  (y+b)^{1-a\over 2}\,,
\end{equation}
and
\begin{equation}
\phi=\phi_\infty+
{k\over 8b}\ln \left( {y-b\over y+b}\right)\,.
\label{keypp}
\end{equation}
In the limit $\mu\to 0$, we have $a\to 0$ and
$b\to \pm |k|/(2\sqrt{6})$, reducing to the solution found in
\cite{Gubser:1999pk,Kehagias:1999tr,Nojiri:1999sb}.

\subsection{The causality bound}
\label{sec4.1}

The causality bound in this case is considerably simpler than that in
the global deformation.
Again we compare two paths connecting two boundary points P and
Q, one along the boundary and the other through the bulk.
The bulk path does not have to be geodesic.
Then the holographic causality condition requires simply that
$hg/r^2 \le 1$.
Since
\begin{equation}
hg/r^2= \left({y-b\over y+b}\right)^a \,,
\end{equation}
the causality restricts $b \ge 0$, which in turn implies
$\mu \ge 0$. Later we shall show that this corresponds to the
requirement of the positive semi-definite mass density.
Hence from now on we consider only the case of positive $\mu$.

\subsection{The repulson}
\label{sec4.2}

As in the case of the global deformation, we probe the
singularity by the massive particle.
We will again find the condition for the repulsion which turns out to
be the same as that of the global deformation.
The spherically symmetric geodesics amount again to a particle motion
in the potential
\begin{equation}
 V(r)={ g\over h e^{\phi\over 2}}
\left(1- {C^2\over ghe^{\phi\over 2}
}\right)\,.
\end{equation}
Note that
\begin{equation}
U_1\equiv ghe^{\phi\over 2}= e^{{1\over 2}\phi_\infty}
{(y-b)^{{k\over 16 b}+{3a+1\over 4}}
\over   (y+b)^{{k\over 16 b}+{3a-1\over 4}}
}\ ,
\end{equation}
and
\begin{equation}
U_2\equiv
{he^{\phi\over 2}\over g}= {e^{{1\over 2}\phi_\infty}\over (y-ab)^2}
{(y-b)^{{k\over 16 b}+{3+a\over 4}}
\over   (y+b)^{{k\over 16 b}+{3-a\over 4}}
}\ .
\end{equation}
We are interested in the behavior of the potential as $r\rightarrow
0$. This corresponds to the limit
$y\rightarrow b$.
Since ${k\over 16 b}+{3+a\over 4} > 0$ always holds, $U_2 \rightarrow
0^+$ as $r\rightarrow 0$. Furthermore,
if
\begin{equation}
{k\over 16 b}+{3a+1\over 4} < 0
\ ,
\label{repul}
\end{equation}
then $U_1\rightarrow \infty$, and the potential will have
the hard-core repulsion near the singularity at $r=0$. Otherwise,
$U_1\rightarrow 0^+$, and we will have the infinitely
attractive singularity at $r=0$.

The condition (\ref{repul}) is solved as
\begin{equation}
k < -2(9+\sqrt{57})\mu
\ .
\label{repconp}
\end{equation}
Let us now work out the asymptotic behavior of
the solution. As $r\rightarrow 0$,
\begin{equation}
\psi_0\equiv \psi(0)= {|\mu|\over 2}\left(\sqrt{1+{k^2\over 6\mu^2}}
\,\,\,-\,\,1\right)\ .
\label{psi}
\end{equation}
Using this relation and (\ref{repconp}), one can show that the repulsion
condition is expressed as
\begin{equation}
-{1\over 2}(3+\sqrt{57}) (\simeq -5.27)\
<\ {k\over \psi_0}\ \ \ \ \ ({\rm with}\  k <0)\ ,
\label{psi1}
\end{equation}
which is in agreement with (\ref{repcon}) of the global deformation.
In addition, it is straightforward to show that
\begin{equation}
h\ \rightarrow\ 1+ O(r^{-8})\ ,\ \  \  g/r^2
\ \rightarrow\ 1-\mu r^{-4} + O(r^{-8})
\ ,\ \ \  \phi
\ \rightarrow\ \phi_\infty-{k\over 4} r^{-4} + O(r^{-8})\ ,
\end{equation}
as $r\rightarrow \infty$.
Incidentally, from the correspondence of (\ref{expec}),
the negative $k$ corresponds to the magnetic background,
as noted in the end of Section \ref{sec2}.

\subsection{The ADM mass}
\label{sec4.3}

The parameter $\mu$ can be interpreted as the mass parameter of the
geometry like $A$ in the global deformation.
Using (\ref{masses}), we identify the mass of the system as
\begin{equation}
M={3\mu\over 16\pi G}\int d\vec{x} \ .
\end{equation}
Therefore the mass density of the boundary CFT is given by
\begin{equation}
T^{bd}_{00}={3 \mu\over 16\pi G}={3 N^2\mu\over 8\pi^2 l^4}\ ,
\end{equation}
where we have restored $l$ dependence with the dimensionless
quantity $\mu$
in the unit of $l$.

In conclusion, for the  Poincar\'e patch solution the
causality implies just the exclusion of the negative mass density. For
the causal case, we expect to have in general two phases depending on the
expectation value of ${\cal L}_{CFT}$.
When $k < -2(9+\sqrt{57})\mu$, the theory is confining,
as we will show in Section \ref{sec6}.

\section{Nature of Singularity}
\label{sec5}

In this section we discuss the nature of singularity in the
Poincar\'e deformation.
In the global deformation, we have found that the singularity
is timelike. In the present case, one finds that $hg=0$ and $h/g=0$ at
the singularity $y=b$.
Thus at first glance the singularity might appear lightlike.
However, as pointed out in \cite{Hellerman}, this conclusion could
be premature and merely an artifact of the particular coordinate system.
We now show that this is indeed the case. Note that the
metric in fact can be represented in terms of $y$ coordinate by
\begin{equation}
ds^2= (y-b)^{1-a\over 4}(y+b)^{1+a\over 4}
\left(
- \left({y-b\over y+b}\right)^a dt^2 + {dy^2\over 16
(y-b)^{5-a\over 4}(y+b)^{5+a\over 4}}
 + d\vec{x}^2
\right)
+ d\Omega^2_5\ .
\label{poinm}
\end{equation}
To see what is happening in the Penrose diagram in $(t,y)$-plane,
let us introduce the coordinate $w$ defined by
\begin{equation}
w(y)= \int^\infty_y {d\tilde{y}\over
(\tilde{y}-b)^{5+3a\over 8}(\tilde{y}+b)^{5-3a\over 8}
}
\ .
\end{equation}
The range of $w$ is given by $[0,\,\, w_0]$
with $w_0=w(b)\,\,>\,\,0$ for $k\neq 0$. In terms of this variable,
the metric with fixed $\vec{x}$ and the angular coordinates becomes
\begin{equation}
ds^2= (y-b)^{1+3a\over 4}(y+b)^{1-3a\over 4}
\, d(w-t) d(w+t)\ .
\end{equation}
Introducing further $u_\pm= \tan^{-1}(w\pm t)$, the metric becomes
\begin{equation}
ds^2= (y-b)^{1+3a\over 4}(y+b)^{1-3a\over 4}\sec^{2}u_+\sec^{2}u_-
\,\,du_+ d u_-\ .
\end{equation}
The range of $u_\pm$ is given by $[-{\pi\over 2},{\pi\over2}]$
with $0 \le u_++u_-$ and $\tan u_++\tan u_-\le 2 w_0$. The Penrose
diagram is depicted in Figure 5.
\begin{figure}[htb]
\epsfxsize=2in
\centerline{
\epsffile{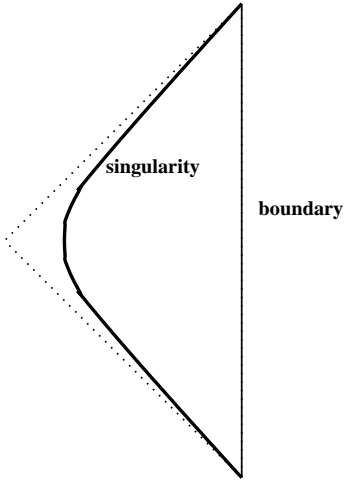}
}
\vspace{.1in}
\caption{\small The Penrose diagram for the  Poincar\'e deformation.}
\end{figure}
Hence it is now clear that the trajectory of singularity is timelike.

Also for any finite size box in the direction of $\vec{x}$, the singularity
is essentially pointlike due to the scale factor $r^2$ that
becomes zero as one approaches the singularity.

\section{Wilson Loop and Confinement}
\label{sec6}

We argued in Section \ref{sec3.4} that the infinitely strong repulsion
is indicative of the confinement in the dual gauge theory. Our
argument was based on an extrapolation of the behavior of the static
string from that of the massive particle in the global deformation.
In this section, exploiting the exact analytic form of the solution in
the Poincar\'e patch, we will make our claim precise, by directly
calculating the static string corresponding to the Wilson loop.

The action for the fundamental string is given by
\begin{equation}
S= -{1\over 2\pi \alpha'} \int d\tau d\sigma
\sqrt{ -\mbox{det}(g_{\mu\nu} \partial_a X^\mu\partial_b X^\nu)}\ ,
\end{equation}
where $g_{\mu\nu}$ is the string frame metric. We assume that the Wilson
loop is static and choose the gauge $\tau=t$ and $\sigma=y$.
We shall consider the case where the Wilson loop is independent of
$x_2$ and $x_3$. The Nambu-Goto Lagrangian becomes
\begin{equation}
L= - \int dy \sqrt{ A(y)\left(B(y)+ C(y)\left(
{dx/ dy}
\right)^2\right)}\ ,
\end{equation}
where we are considering the string frame metric of the form
\begin{equation}
ds^2= - A(y) dt^2 + B(y) dy^2 +C(y) dx^2\ ,
\end{equation}
which is found from (\ref{poinm}) and (\ref{keypp}) as
\begin{eqnarray}
A(y) &=& e^{\phi_\infty/2} (y-b)^{\frac{1+3a}{4}+\frac{k}{16b}}
(y+b)^{\frac{1-3a}{4}-\frac{k}{16b}}\ , \nonumber\\
B(y) &=& \frac{1}{16} e^{\phi_\infty/2} (y-b)^{-1+\frac{k}{16b}}
(y+b)^{-1-\frac{k}{16b}}\ , \\
C(y) &=& e^{\phi_\infty/2} (y-b)^{\frac{1-a}{4}+\frac{k}{16b}}
(y+b)^{\frac{1+a}{4}-\frac{k}{16b}}\ . \nonumber
\end{eqnarray}
The equation of motion is given by
\begin{equation}
{d\over dy}\left({ \sqrt{A}\, C\, {dx/dy} \over \sqrt{ B+ C\left(
{dx/ dy}
\right)^2}
} \right)=0\ .
\end{equation}
Integrating this once, we get
\begin{equation}
{ \sqrt{A}\, C\, {dx/dy} \over \sqrt{ B+ C\left(
{dx/ dy}
\right)^2}
}=\pm q^{-2}\ .
\label{velo}
\end{equation}
A further integration leads to the solution
\begin{equation}
x-x_0=\pm \int dy
{ \sqrt{B}\over \sqrt{C} \sqrt{q^4 AC-1 }
}\ .\label{timex}
\end{equation}
To understand what it implies, let us rewrite (\ref{velo}) in the form
\begin{equation}
\left({dy\over dx}\right)^2+ {\cal V}(y)=0\ ,
\end{equation}
with the potential
\begin{equation}
{\cal V}(y)={C\over B} (1-q^4 AC)
=16{(y-b)^{\ 5-a\over 4} \over (y+b)^{\!-\!5-a\over 4}}
\left(1-e^{\phi_\infty}q^4{(y-b)^{{\ 1+a\over 2}+{k\over 8b}}\over
(y+b)^{{\!-\!1+a\over 2}+{k\over 8b}}}
\right)
\ .
\end{equation}
This can be viewed as a zero-energy particle moving in one dimension
under the potential ${\cal V}$,
regarding the coordinate $x$ as the `time'.

The confinement will occur when the zero-energy `particle' spends
an arbitrarily large `time', when it approaches the
turning point denoted by $y_0$. At the turning point, one has $dy/dx=0$.
Thus ${\cal V}(y_0)=0$, which implies that
\begin{equation}
q_0^4 A(y_0)C(y_0)=1\ ,
\end{equation}
for an appropriate choice of the integration constant $q=q_0$.
The condition of arbitrarily large `time' spending
may be fulfilled if ${\cal V'}(y_0)=0$. This leads to
\begin{equation}
(\ln A C)'|_{y=y_0}={y_0 + ab+ k/4\over y_0^2-b^2}= 0\ ,
\end{equation}
where the condition ${\cal V}(y_0)=0$ is used. For the existence of
the solution in the range $y \in (b,\infty)$, one has to satisfy
\begin{equation}
y_0-b = -ab- k/4-b \equiv 2b \beta> 0\ .
\end{equation}
because $y > b$. The solution is given by
\begin{equation}
k < -12 \mu \ .
\label{wilsoncon}
\end{equation}
Then ${\cal V}(y_0)=0$ is satisfied by choosing the
integration constant $q$ as
\begin{equation}
q_0^4= {1\over 2b}\beta^\beta (1+\beta)^{-(1+\beta)}
e^{-\phi_\infty}\ .
\end{equation}
If these requirements are met, the potential may be approximated by
\begin{equation}
{\cal V}= - \kappa^2 (y-y_0)^2 \,+ \cdots\ ,
\end{equation}
near the turning point for some constant $\kappa$. The solution of
(\ref{timex}) is given in the form of $x-x_0 \simeq \pm
{1\over\kappa}\ln |y-y_0|$, which is consistent with
the condition of arbitrarily large time spent by the particle
approaching $y_0$.

Let us now describe how the string behaves
depending on various integration constants and parameters.

\subsection{Coulombic case}
\label{sec6.1}

For $\beta < 0$ (or equivalently $k > -12 \mu$)
and any $q$, the potential starts from large negative values for
large $y$, crosses zero at $y=y_0$, turns around at some $y$
($<y_0$) and approaches zero as $y\rightarrow b$, as depicted
in~Figure~\ref{pot1}.

\begin{figure}[htb]
\epsfxsize=2.4in
\centerline{
\epsffile{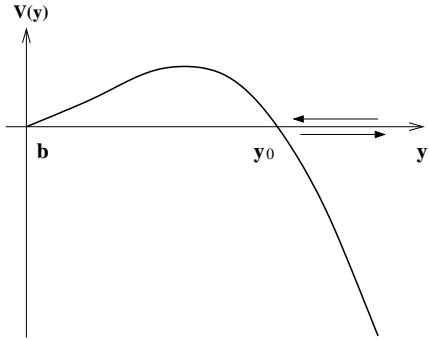}
}
\vspace{.1in}
\caption{\small The shape of the potential for $\beta < 0$.}
\label{pot1}
\end{figure}

The zero-energy particle turns around at $y=y_0$ without spending
much `time'. The behavior of the string is not much different from
the pure AdS case where the quark-antiquark potential
is Coulombic.  Therefore this case corresponds to the Coulomb phase.

For $\beta >0$ and $q <q_0$, the shape of the potential is again
similar to the pure AdS case. In particular for $q \ll q_0$,
the shape of the string and potential is not much different from those
in the pure AdS. The string remains in the asymptotic region where
the geometry approaches the pure AdS spacetime (see
Figure~\ref{pot2}).
For small $q$, the separation between the quark and antiquark is of
the order of $q$ according to the IR/UV relation.
The energy scale here is much higher than that of the
confinement. Thus the quark-antiquark potential is essentially
Coulombic.

\begin{figure}[htb]
\epsfxsize=3.0in
\centerline{
\epsffile{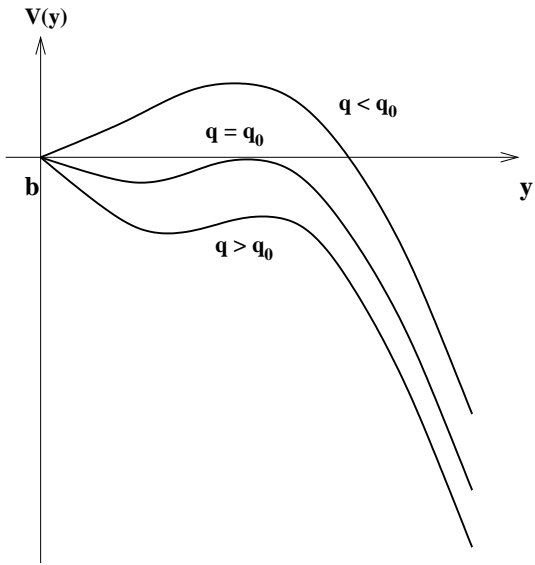}
}
\vspace{.1in}
\caption{\small The shape of the potential for $\beta > 0$.}
\label{pot2}
\end{figure}

\subsection{Confining case}
\label{sec6.2}

When $\beta>0$ and $q$ approaches $q_0$ from below, the string spends
more and more \lq time' near the turning point $y\sim y_0$.
The separation of the quark and antiquark becomes larger and larger
when one sends $q$ to $q_0$ from below because the `time' spent near
the turning point increases more and more, as depicted in Figure~\ref{wl}.

\begin{figure}[htb]
\vskip .5cm
\epsfxsize=3.6in
\centerline{
\epsffile{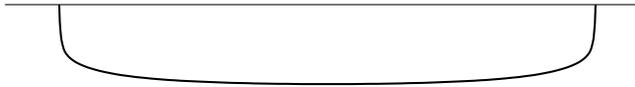}
}
\vspace{.1in}
\caption{\small A string configuration representing confinement.}
\label{wl}
\end{figure}
In the limit $q \to q_0$, we can compute the tension of the QCD string and
the energy scale of the confinement. The energy of the string is given by
\begin{equation}
E=  \int dy \sqrt{ A\left(B+ C\left(
{dx/ dy}
\right)^2\right)}
=\int dx \sqrt{q^4 A^2 C^2}\ ,
\label{qenergy}
\end{equation}
where we have used the equation of motion. The integral in fact diverges
and one may regulate it by subtracting the self-energy of quark and antiquark,
which is twice of the quark self-energy obtained by setting $x=0$:
\begin{equation}
E_q=  \int_b^\infty dy \sqrt{AB}
\ .
\end{equation}
Then the quark-antiquark potential is given by
\begin{equation}
V_{q\bar{q}}= 2\int_{y_0}^\infty dy \sqrt{AB}\;\sqrt{\frac{q^4 AC}{q^4 AC-1}}
-2\int_b^\infty dy\sqrt{AB}
\ .
\end{equation}

Since  $q_0^4 A(y_0) C(y_0)=1$ and  the string stays
near the turning point
for most of the \lq time', we find from (\ref{qenergy}) the tension of
the QCD string to be
\begin{equation}
T_{QCD}= \sqrt{A(y_0)C(y_0)}=q_0^{-2}=
\sqrt{\mu} {(1+\beta)^{1+\beta\over 2}
\over \sqrt{a}\,\, \beta^{\beta\over 2}}e^{\phi_\infty/2}\ ,
\end{equation}
with $\Delta E= T_{QCD}\Delta x$. This sets the scale of
the confinement, and was previously calculated in \cite{Gubser:1999pk}
in the $\mu\to 0$ limit.
When the separation $L$ is much larger than
$q_0=1/\sqrt{T_{QCD}}$, we are in the confining phase.

When the integration constant $q$ is larger than $q_0$ with $\beta >0$,
one finds that the turning point corresponds to $y=b$ and
the string touches the singularity at the turning.
However the string coupling becomes large near $y=b$ so that we cannot
make a definite statement as to what is physically happening in this
regime. Note also that the regular branch $q \le q_0$ alone covers all
the energy scale of the boundary field theory.

Finally if one rewrites the condition for the confinement
of (\ref{wilsoncon}) in terms of ${k\over \psi_0}$ using
(\ref{psi}), one gets a weaker condition,
 \begin{equation}
-6  \
<\ {k\over \psi_0}\ \ \ \ \ ({\rm with}\  k <0)\ ,
\end{equation}
than the one in (\ref{psi1}). Since this condition is by nature
local in the bulk geometry, one may expect that the same may be true
for the global deformation.

\subsection{The Mass Gap}
\label{sec6.3}

As further evidence for the confinement, we will show there exists a
mass gap in the dual gauge theory, following \cite{Witten:1998zw}.
The same calculation was done in \cite{Gubser:1999pk} for
$\mu=0$. Here we will generalize their result to the case of
non-vanishing $\mu$ ($>0$).

We consider the fluctuation $\delta\phi$ of the dilaton about our
background, and see if $\delta\phi$ has the discrete spectrum with a
mass gap.
To show the mass gap, it is sufficient to think of the s-wave on the
five sphere, thus setting $\delta\phi=\varphi(z)e^{i\omega t}$.
Then the fluctuation mode obeys
\be
{d^2\over dz^2}\varphi + V(z)\varphi = 0\ ,
\ee
with
\be
V={\omega^2\over 16} {e^{{3\over 2}abz}\over \sinh^{3\over 2} bz}\ .
\ee
The coordinate variable $z$ is related to $y$ by
\be
z=-\int {dy\over y^2-b^2}={1\over 2b}
\ln \left({y+b\over y-b}\right)\ .
\ee
The existence of the gap for $k/\mu=-\infty$ ($\mu\to 0$), was
shown in \cite{Gubser:1999pk}, and a few lowest values of
$\omega^2=m^2l^2$ for $a=0$ are given in the Table 1 of their paper.

We have carried out a numerical analysis which shows the mass gap for
$\mu> 0$ with some upper bound on $\mu$, as anticipated
from the condition (\ref{wilsoncon}) for the confinement.  To
illustrate this we plot the mass of the lowest mass $(l=0)$ glueball
as a function of $a$ in units where  $b=1$. The parameters $a,b$ are
related to $k, \mu$ by (\ref{abrelation}). Note that the glueballs
cease to exist for $a=1$, indicating no confinement.

\begin{figure}[htb]
\vskip .5cm
\epsfxsize=2.8in
\centerline{
\epsffile{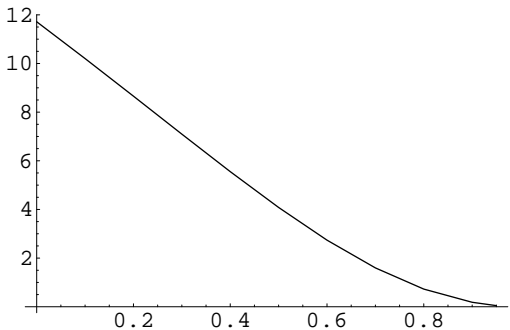}
}
\vspace{.1in}
\caption{Mass of lowest mass glueball as function of $a$.}
\label{glueb1}
\end{figure}

\section{Discussions}

In this paper, we have
proposed gravity duals of certain non-supersymmetric backgrounds or
states in ${\cal N}=4$ SYM which give the non-vanishing expectation
values to the dimension four operators, ${\cal L}_{CFT}$ and $T_{00}$.
The backgrounds are homogeneous and time-independent, and respect the
full $SO(6)$ R-symmetry.
Our proposed gravity duals are of non-supersymmetric dilatonic
deformation. In particular we have studied two cases in detail -- the
global and Poincar\'e deformations, whose dual field theory lives on
$R\times S^3$ and $3+1$ dimensional Minkowski spacetime respectively.

We have carried out numerical studies, for the most part, in the case of
the global deformation, whereas we have found the exact analytic
solution for the Poincar\'e deformation.
It is shown that the naked singularity is inevitable for this
class of deformations and it is in fact timelike.
Then the causality bound of \cite{Kleban:2001nh} is used as a
criterion to determine which of our singular geometries may make sense
in string theory. This leads us to the positive lower bound $A(k)$ on
mass $A$ for a given $k$ in the global deformation, whereas to the
positivity of mass $\mu$ (irrespective of $k$) in the Poincar\'e deformation.
The causality bound is predominantly determined by the contribution
from the spacetime region far off the singularity. Thus our estimate
of the causality bound should be plausible, although the supergravity
approximation breaks down near the singularity.
It is our claim that the singular geometries we have found are sensible in
the AdS/CFT correspondence, as long as the causality bound is
satisfied, irrespective of the validity of our approximation, while
the accuracy of our analysis is only reliable away from the singularity.

Another criterion for a solution to be physical is given in
Ref.~\cite{Gubser:2000nd}, which states that large curvature naked
singularities are allowed only if the scalar potential is bounded from
above in the solution. If we accept this criterion, our solution
is physical irrespective of whether $\mu=0$ or not since the scalar
potential is zero identically.


Our proposed criterion for the sensibility of geometry with a
naked singularity crucially relies
on the existence of the well-defined dual field theory state.
Also there is no credible way to argue the stability of our geometry
within the supergravity approximation. The question of stability can
thus be checked only through the confirmation of the existence of
the field theory state and study of its properties.

Here we give a brief argument to bolster the existence of such a state.
Let us first consider the case $\mu=0$. In this case the Lorentz
invariance is restored, as can be seen, for example, from
the metric (\ref{poinm}) with $a=0$. This is consistent with
the dual field theory description where $\langle T_{00}\rangle=0$
and $\langle T_{0i}\rangle=\langle T_{ij}\rangle=0$.
Also having the expectation value for the Lagrangian density,
$\langle{\cal L}\rangle=k/4$, respects the Lorentz invariance.
Since $T_{\mu\nu}=F_{\mu\alpha}F_{\nu}^{\hspace{0.1cm}\alpha}
-{1\over 4}\eta_{\mu\nu}F^2$ for $X^{I=1\cdots 6}=0$,
these imply that
$-\langle F_{0\alpha}F_0^{\hspace{0.1cm}\alpha}\rangle
=\langle F_{i\alpha}F_i^{\hspace{0.1cm}\alpha}\rangle
={1\over 4}\langle{\cal L}\rangle=k/16$ and
$\langle F_{0\alpha}F_i^{\hspace{0.1cm}\alpha}\rangle
=\langle F_{i\alpha}F_j^{\hspace{0.1cm}\alpha}\rangle
=0$ for $i\ne j$.
Such a state can exist, though its explicit construction is beyond
our scope.
Raising the value of $\mu$ breaks the Lorentz invariance but keeps the
$SO(3)$ rotation symmetry. In fact it yields
$\langle T_{00}\rangle={3N^2\mu\over 8\pi^2}$ and
$\langle T_{0i}\rangle=\langle T_{ij}\rangle=0$
in the dual field theory, matching the symmetry on both sides.
By a similar argument, again we do not see any contradiction for the
existence of such a state.

It seems also suggesting that our geometry is stable, since the dual
field theory state is an energy eigenstate and thus cannot decay into
anything else by the unitary evolution driven by the ${\cal N}=4$ SYM
Hamiltonian.


As a physical implication,
we expect that the confinement may occur in our
non-supersymmetric dilatonic deformation, since the deformation
introduces a mass scale in the theory. We have argued that, for the
confinement to actually happen, it would be signaled by the hard-core
repulsion from the naked singularity for one possibility.
We have shown that there is a certain range of parameters for the
global/Poincar\'e deformations in which the geometry indeed
exhibits the strong repulsion. Both in the global and Poincar\'e
deformations, we find an overlap between the ranges of parameters for
the repulsion and causality.

However, it turns out that the would-be confining scale in the global
case is too large compared to the size $l$ of the system.
Therefore the confinement representing a linear potential between
quark and antiquark as its defining property,
cannot be seen  for the global deformation.
On the other hand, in the  Poincar\'e deformation,
it is shown that the confinement indeed occurs, for a certain range of
parameters, by the explicit calculation of the
static string corresponding to the Wilson loop.
The confinement in this case is in fact a consequence of the strong
repulsion from the singularity as depicted in Figure \ref{wl}.

What would be the objects located at the origin of AdS space in the
global deformation? Clearly these objects
are zero dimensional, which are distributed over $S^5$ uniformly. They
produce no R-R flux and only accompany the non-vanishing dilaton.
Furthermore these objects are non-supersymmetric. There are only
two candidates in the IIB closed string theory with these specifications.
One is the unstable D0-brane and the other is the AdS-Schwarzschild
black hole.  Because our solution does not involve any horizon,
we may exclude the possibility of the black hole. Thus it is tempting
to identify the timelike naked singularity with unstable D0-branes,
sitting at the center of AdS space.

The above solutions involve three free parameters, two of which are
physically relevant.
The parameter $k$, the strength of the dilaton, would be
related with the number of D0-branes.
The parameter $A$, the ADM mass, should represent the total
eigen-energy of the collection of unstable D0-branes
(see \cite{Dru,Harvey:2000qu} for related discussions).

It is interesting to try to understand the physics of the unstable
D0-branes more deeply. The analysis of the rolling tachyon boundary CFT
\cite{Sen:2002nu} for the decay of an unstable brane~\cite{Sen:2002in}
suggests that the D0-branes decay into pressureless dust.
The energy-momentum tensor $T_{00}= {\tau_0\over 2}(\cos 2\pi \lambda +1)$
depends on a parameter $\lambda$ of the boundary CFT. A particular simple
interpretation of the tachyon matter was suggested in
\cite{Lambert:2003zr}, where it was argued that the unstable brane
decays primarily into closed string modes which (at weak coupling) are
long lived. Furthermore the massive string states that have
nonrelativistic velocities are localized near the locus of the D0-brane
for a long time.  Note that
for $k>0$ the string coupling becomes arbitrarily weak at the location
of the singularity and one might expect that the tachyon matter does
not diffuse at all. There is also a new feature in AdS, where massive
states are confined by the AdS curvature from propagating away from
the center of AdS. It would be interesting to make this stringy
resolution of the singularity more precise. Unfortunately the exact
boundary states are only known in flat space and not available for the
$AdS_5\times S^5$ background. With present techniques, only an analysis
using the tachyon effective action coupled to type IIB supergravity in
the spirit of \cite{Buchel:2002tj,Leblond:2003db} seems feasible. We
will leave this interesting question for future work.

In particular we do not know the exact relation of the parameters $A$
and $k$ of the supergravity solution to the parameters of the D0-branes.
It is suggestive to interpret the marginal case, $A(k)$, will correspond
to the unstable D0-branes in their ground state.

The decay process of the tachyon could be described by an S-brane type
of solutions~\cite{Sbrane,Sbrane2,Sbrane3,Sbrane4}. In the past the
formation of naked singularities in numerical relativity using finely
tuned incoming dilaton and graviton excitations has been studied in detail
\cite{Choptuik:jv}. A possible scenario is to consider an incoming pulse
of dilaton and graviton which forms a naked singularity is actually an
S-brane process of the formation of an unstable D0-brane. The solution
should correspond to time-dependent space-like brane in AdS space.
One possible line of further investigation is to study such processes
numerically (see \cite{Pretorius:2000yu,Husain:2002nk} for
first steps in this direction). It would also be
interesting to find the corresponding solutions and
study their relevance to the present problem.

In conclusion we have found a very simple example of the confinement
of YM theories via AdS/CFT. The bulk mechanism behind the confinement
is speculated due to the presence of the unstable D0-branes.

\section*{Acknowledgement}

DB is supported in part by KOSEF ABRL R14-2003-012-01002-0 and KOSEF
R01-2003-000-10319-0.
The work of MG was supported by NSF grant
0245096. Any opinions, findings and conclusions expressed in this
material are those of the authors and do not necessarily reflect the
views of the National Science Foundation.
The work of SH was supported in part by the Israel Science
Foundation under grant no. 101/01-1, and
that of NO was supported in part by Grants-in-Aid for Scientific Research
Nos. 12640270 and 02041.

\appendix

\section{Hyperbolic Slicing}

In sections \ref{sec3} and \ref{sec4}, we study in detail the
dilatonic deformations Eqs.(\ref{metric})-(\ref{5form}) in the global
coordinates and the Poincar\'e patch respectively.
In this appendix, we briefly discuss the case of $K=-6$, i.e. the
hyperbolic slicing.

Our main equation (\ref{key}) in this case is given by
\begin{equation}
r (\ln \psi')'= {6 r^2 -1\over 2 r^2 -1} + {k^2\over 6\psi^2}\ .
\label{keyh}
\end{equation}
To see what is happening, let us look at the case where $k=0$.
The solution reads
\begin{equation}
g=r^2-1 -A/r^2\,, \ \ \ \ \ \  h=1\,.
\end{equation}
The claim is that the singularity is a three-dimensional hypersurface
which is extended to the boundary.

To show it, we choose the metric of $3$-dimensional hyperboloid to be
\begin{equation}
ds_{3,K=-6}^2=
{1\over \tilde{w}^2}(d\tilde{w}^2+dx_1^2+dx_2^2)
\ .
\end{equation}
In the case where $A=0$, introducing new coordinates
\begin{equation}
\tilde{t}=\tilde{w}\sin \theta \sinh t\,,\ \
\tilde{x}_3=\tilde{w}\sin \theta \cosh t\,,\ \
\tilde{z}= \tilde{w}\cos \theta\ ,
\end{equation}
with $r=1/\cos\theta$, one finds that
the metric (\ref{metric}) is
\begin{equation}
ds^2= {1\over \tilde{z}^2}
\left(- d\tilde{t}^2 +  d\tilde{z}^2 + dx_1^2+ dx_2^2+
d\tilde{x}_3^2\right)
+ d\Omega^2_5\ ,
\label{poincare}
\end{equation}
i.e. the same as that in the Poincar\'e patch.
When $A$ is non-vanishing, the event horizon appears at the fixed
$\theta=\theta_0$, determined by
\begin{equation}
K(\theta)=r^2-1-A/r^2=\tan^2\theta-A\cos^2\theta=0\,.
\end{equation}
For $t=0$ with $\theta=\theta_0$, it describes a three dimensional
surface extended along $x_1$, $x_2$ and $(\tilde{z},\tilde{x}_3)=\tilde{w}
(\cos\theta_0,\,\sin\theta_0)$. For $t\ne 0$, one finds that
the shape is time dependent in the standard coordinates of the
Poincar\'e patch in (\ref{poincare}).

\end{document}